# A New Deep Learning Method for Image Deblurring in Optical Microscopic Systems

Huangxuan Zhao [a,c,d,1], Ziwen Ke [b,f,1], Ningbo Chen [a,1], Ke Li [c,d], Lidai Wang [e], Xiaojing Gong [a], Wei Zheng [a], Liang Song [a], Zhicheng Liu [c,d,*], Dong Liang [b,*], Chengbo Liu [a,*]

[a] *Research Laboratory for Biomedical Optics and Molecular Imaging, Shenzhen Institutes of Advanced Technology, Chinese Academy of Sciences, Shenzhen 518055, China*

[b] *Research Center for Medical AI, CAS Key Laboratory of Health Informatics, Shenzhen Institutes of Advanced Technology, Chinese Academy of Sciences, Shenzhen 518055, China*

[c] *Beijing Key Laboratory of Fundamental Research on Biomechanics in Clinical Application, Beijing 100069, China*

[d] *School of Biomedical Engineering, Capital Medical University, Beijing 100069, China*

[e] *Department of Mechanical and Biomedical Engineering, City University of Hong Kong, 83 Tat Chee Ave, Kowloon, Hong Kong SAR, China*

[f] *Shenzhen College of Advanced Technology, University of Chinese Academy of Sciences, Shenzhen 518055, China*

**Abstract**

Deconvolution is the most commonly used image processing method to remove the blur caused by the point-spread-function (PSF) in optical imaging systems. While this method has been successful in deblurring, it suffers from several disadvantages including being slow, since it takes many iterations, suboptimal, in cases where experimental operator chosen to represent PSF is not optimal. In this paper, we are proposing a deep-learning-based deblurring method applicable to optical microscopic imaging systems. We tested the proposed method in database data, simulated data, and experimental data (include 2D optical microscopic data and 3D photoacoustic microscopic data), all of which showed much improved deblurred results compared to deconvolution. To quantify the improved performance, we compared our results against several deconvolution methods. Our results are better than conventional techniques and do not require multiple iterations or pre-determined experimental operator. Our method has the advantages of simple operation, short time to compute, good deblur results and wide application in all types of optical microscopic imaging systems. The deep learning approach opens up a new path for deblurring and can be applied in various biomedical imaging fields.

*Abbreviations:* PSF, point-spread-function; CNN, Convolutional Neural Network; OR-PAM, optical-resolution photoacoustic microscopy; OCT, Optical Coherence Tomography; MSE, mean squared error; PSNR, peak signal to noise ratio; SSIM, Structural Similarity.

---

[*] Corresponding authors:
  E-mail addresses: zcliu@ccmu.edu.cn (Z. Liu), dong.liang@siat.ac.cn (D. Liang) and cb.liu@siat.ac.cn (C. Liu)
  [1] These authors contributed equally to this work.

## 1. INTRODUCTION

Optical microscopic imaging systems are widely used to analyze structures and functions of living biological cells, tissues, and organs in biomedical applications [1-3]. However, due to the non-ideal PSF of the system, they always blur the imaging objects. The blurring of an object is mathematically described as a convolution operation between object and PSF of the imaging system [4, 5]. If the blurring is not addressed, identifying and analyzing objects in the images would be very difficult in many cases [4, 6, 7]. A deconvolution concept, which tries an inverse process of the convolution, has become an indispensable part of image and signal processing to deblur the images. Thus, deconvolution is widely used in imaging systems to enhance the images and to deblur the imaging objects. This method is also used in all types of optical microscopic imaging systems, such as optical microscope, confocal microscope and photoacoustic microscope [8-12].

The deconvolution operation attempts to find the optimal solution through linear or nonlinear equations [12, 13]. Two popular deconvolution methods most commonly used are Richardson-Lucy's (RL) and statistical blind deconvolution (Deconv) [10, 14, 15]. These two methods are robust and reconstruct high-quality images even in the presence of noise. However, despite improvements made to the deconvolution method over the years, there is still room to improve deblurring outcomes due to the following limitations of deconvolution: (1) It is often difficult to determine the optimal PSF of the system experimentally, leading to suboptimal deblurring [4, 16]; (2) Deconvolution is computationally expensive due to the inverse operation and several numbers of iterations involved, for examples, it takes tens of seconds to process a simple 2D image by using deconvolution method as used in [8, 9]. Thus improvements in deblurring are needed in terms of performance and independence from the operator. New deblurring approaches are continuously being investigated to address the above limitations [17-19]. In this paper, we are proposing a novel deep-learning-based deblurring method applicable to all optical microscopic imaging systems.

Deep learning is a class of machine learning techniques that use multilayered artificial neural networks for the automated analysis of signals or data [20]. Convolutional Neural Network (CNN), composed of a convolution layer and a nonlinear operator, is a popular embodiment of deep learning technique. The CNN fits nonlinear equations by machine learning rather than manually providing equations. The fitting results using CNN have exceeded the performance of many traditional nonlinear algorithms [21-24]. The super-resolution technology developed based on CNN improves blurring by up sampling the low-resolution images [24-26]. Inspired by the super-resolution technology, researchers used a low magnification objective as input and a high magnification objective as ground-truth for training CNN to deblur low-resolution images [27, 28]. These low-resolution objectives and high magnification objectives data are generated solely for training. While this super-resolution approach solved the image blurring in optical microscopy to some extent, several drawbacks exist: (1) It cannot be used in imaging systems that do not have a switchable high and low objective lens; (2) The cost of training the network is very high since tens of thousands of extraneous experimental data are required for training purpose only. In recent years, researchers have proposed some new deep learning methods for super-resolution confocal microscopy [29, 30]. However, they tend to have special requirements for imaging equipment (eg., method of [29] was based on photoactivated localization microscopy (PALM), and method of [30] was based on stochastic optical reconstruction microscopy (STORM)). It is difficult to generalize them to various microscopic imaging systems.

Our deep-learning-based deblurring approach does not need any up-sampling (or down-sampling) operation, uses no extraneous experimental data during training processes to train CNN models, and applicable to various conventional microscopes. It only needs to put the data to be reconstructed into the well trained network to generate the deblurred result. We tested our proposed method on database data, simulated data, and experimental data, all of which resulted in good deblurred results. We also quantified the performance improvements and compared them against convolutional deconvolution methods. In the next few sections, we describe our method, present our results, discussion and conclusion.

## 2. METHODS

### 2.1. The main framework of our deblurring method

The main framework of our deblurring method is shown in Fig. 1. We downloaded the dataset from the database FIRE (https://www.ics.forth.gr/cvrl/fire/), which contains 268 images of $2912 \times 2912$ pixels obtained using Optical Coherence Tomography (OCT). We cropped image size to $2304 \times 2304$ pixels containing rich information and used 242 for training and 26 images for testing (Test Data). For the training part, we divided each image into smaller sub-images of size $256 \times 256$ pixels as ground-truth. A total of 19,602 sub-images were obtained for training. We blurred sub-images with different Gaussian function and used them as multiple sets of inputs (the detailed description are presented in *Appendix A*) to obtain outputs through CNN. In this study, the residual dense network (RDN, seen in Fig. 2) was employed in our network. The major components of RDN model to extract features efficiently contain: shallow feature extraction, residual dense blocks (RDBs), global feature fusion, and global residual learning. The detailed information of the CNN architecture is presented in *subsection 2.2*. The loss function we used to train the CNN models is the mean square error (MSE) between outputs and ground-truth. It is minimized, and parameters are constantly updated at each training round. Finally, models fitting the inputs to the ground-truth are obtained. The loss curves are shown in the Fig. $B._1$ and the method to address the over-fitting (or under-fitting) issue, routinely encountered in the deep-learning-based approaches, is provided in the *Appendix B*.

Three types of data were used to test the deblurring ability of our method: Test Data (described above), Simulated Data and Experimental Data. To evaluate the applicability of our trained models, we used two sets of simulated data patterns (showed in Figs. 4(a) and (f)) containing resolution target and checkerboard, which are often used in testing the resolution and distortion of reconstructed images. For experimental data, we obtained optical microscope imaging data (2D data) and optical-resolution photoacoustic microscopy (OR-PAM) imaging data (3D data) of the rat eyes in our lab. More details on the experimental data are provided in *subsections 2.3* and *2.4*. All animal handling, and experimental procedures were confirmed to the protocol approved by the Animal Study Committee of Shenzhen Institutes of Advanced Technology, Chinese Academy of Sciences. The simulated patterns are generated by MATLAB software (R2017a, Mathworks, Natick, MA, USA).

To compare outcomes quantitatively, we used the three most commonly used analyses: Mean Squared Error (MSE), Peak Signal to Noise Ratio (PSNR), and Structural Similarity (SSIM). The smaller the MSE, larger the PSNR and SSIM, the more similar the reconstructed images are to the original images. We compared our results against results obtained using other techniques by analyzing these three quantitative parameters for quantifying the improvement.

## 2.2. Convolutional neural network architecture

In deep learning based methods, a deep model is constructed to learn experiences from a large number of training data by an optimization algorithm. The goal of deep learning is to enable the well-trained models to produce appropriate outputs when fed with new and unseen input data. In this study, we propose a new optical image deblurring method, which was employed as a residual dense network (RDN, seen in Fig. 2). Specifically, the RDN model contains five major components: shallow feature extraction, residual dense blocks (RDBs), global feature fusion, global residual learning and up-scaling. The forward process of the entire network can be briefly described as follows.

First, low-resolution optical images $I_{LR}$ are fed into the network for shallow feature extraction. Here, we use two convolutional layers to extract the shallow features:

$$\begin{cases} F_{-1} = \delta(W_0 * I_{LR} + b_{-1}) \\ F_0 = \delta(W_{-1} * I_{-1} + b_0) \end{cases} \quad (1)$$

where $W_{-1}, W_0, b_{-1}, b_0$ are the convolutional filters and biases respectively. $F_{-1}$ and $F_0$ are the shallow features extracted. $\delta$ is the activation function for nonlinearity.

Second, the shallow features go through D (D=4 in this study) RDBs for local feature fusion. The output $F_d$ of the d-th RDB can be expressed in the following equation:

$$F_d = f_{RDB,d}(F_{d-1}) = f_{RDB,d}(f_{RDB,d-1}(...(f_{RDB,1}(F_0))...)) \quad (2)$$

where $f_{RDB,d}$ denotes the forward process of the d-th RDB. Each RDB includes dense connections, local feature fusion, and local residual connections. Dense connections refer to the direct connections of each convolutional layer to the subsequent layers, which can enhance the transmission of local features and makes full use of features from all the preceding layers. All local features are concatenated together and pass through a 1×1 convolutional layer to achieve local feature fusion. Supposing each RDB has C (C=5 in this study) 3×3 convolutional layers, then the local features fusion can be defined as:

$$F_{d,\ LF} = W_d^{1*1}([F_{d-1}, F_{d,1}, ..., F_c, ..., F_{d,C}]) + b_d^{1*1} \quad (3)$$

where $W_d^{1*1}$ and $b_d^{1*1}$ denotes the weights and biases of the 1×1 convolutional layer in the d-th RDB. $[F_{d-1}, F_{d,1}, ..., F_c, ..., F_{d,C}]$ refers to the concatenation of the local features of the d-th RDB. Then residual connections are introduced in RDB to further improve the information propagation:

$$F_d = F_{d-1} + F_{d,\ LF}. \quad (4)$$

Third, these residual dense features from D RDBs are merged via global feature fusion (concatenation + 1×1 convolution):

$$F_{GF} = W_{GF}^{1*1}([F_1,\ F_2, ..., F_D]) + b_{GF}^{1*1}. \quad (5)$$

Fourth, the global residual connection combines the shallow features with the global fused features:

$$F_{GR} = F_{-1} + F_{GF}. \quad (6)$$

Finally, we stack an up-scaling layer in (HR space) to the fine resolution:

$$I_{HR} = \delta(W_{HR} * Upscale(F_{GR}) + b_{HR}) \quad (7)$$

The details of the convolutional layers are shown in Table 1[31, 32]. The mini-batch size was 8. The exponential decay learning rate [33] was used in all CNN-based experiments, and the initial learning rate was set to 0.0001 with the decay of 0.95. All the models were trained by the Adam optimizer [34] with parameters beta_1=0.9, beta_2=0.999 and epsilon=10^-8. We used the mean square error (MSE) between the network outputs and labels as the loss function to train the model.

*2.3. Obtaining optical microscope imaging data*

After sacrificing a healthy rat, the right eye was removed and fixed in 10% buffered formalin (Stephens Scientific, Riverdale, NJ) for 20 hours. A paraffin-cut section method was then used for studying and was submitted for Masson's trichrome staining. Finally, microscopic images were obtained using a Zeiss laboratory light microscope (Axio Lab.A1, ZEISS, Gottingen, Germany) at $10\times$ and $20\times$ magnifications.

*2.4. Obtaining photoacoustic microscope imaging data*

For the photoacoustic imaging, a custom-built OR-PAM system was used to acquire the experimental data of a rat eyeball. The detailed description of the OR-PAM system can be found in our earlier publications [9, 35, 36]. One eight weeks healthy female Sprague Dawley (SD) rat (320 g) was selected for photoacoustic imaging, which remained anesthetized throughout the experiment using 1.5% isoflurane gas (Euthanex, Palmer, Pennsylvania) mixed with oxygen. The right eyelid of the rat was flipped inside out and fixed with adhesive tapes to expose the eyeball and 0.4% oxybuprocaine hydrochloride eye drops were used to anesthetize the eyeball during the imaging process. The imaging head of OR-PAM was positioned directly above the scanning region (i.e., the rat eyeball) during the imaging process.

All three data types were subdivided into a region of $n \times 256 \times 256$ pixels ( e.g., 2D grayscale images were cut into multiple $256 \times 256$ pixels; 2D RGB images were cut into $3 \times 256 \times 256$ pixels; 3D data were cut into $n \times 256 \times 256$ voxels, where n represents the three dimensional numbers of 3D data). All data were input into the fully trained CNN models (the rules for selecting models for input data are presented in *Appendix C*) to get deblurred outputs. Finally, subdivided outputs were stitched together to obtain the final deblurred image. The data were prepared by MATLAB, and the models were implemented in open framework Tensorflow [37] with CUDA and CUDNN support. The model was run on an Ubuntu 16.04 LTS (64-bit) operating system equipped with a Xeon Silver 4110 Central Processing Unit (CPU) and NVIDIA Quadro P4000 Graphics Processing Unit (GPU, 8GB memory).

3. RESULTS AND DISCUSSION

*3.1. Test Data*

We applied different training models that corresponded to the Test Data. For example, when the Test Data was blurred by a Gaussian window with $\sigma=4$, we used the training model trained by the data that was also blurred using a Gaussian window with $\sigma=4$. Compared to the Deconv and RL deconvolution methods (iterations=20, which was an experimental value used by many studies, such as references [8, 9]), our method showed superior results. The results $\sigma=4$ are selected for illustration purpose and are shown in Fig. 3. Figs. 3(a), (b), (c), (d) and (e) are the original image, the Gaussian blurred image with $\sigma=4$, deblurred image using Deconv method, deblurred image with RL method, and deblurred image with our method, respectively. Figs. 3(f), (g), (h), (i), and (j) are the enlarged view of the subareas in

Figs. 3(a), (b), (c), (d) and (e), respectively. The subareas are indicated by the green dotted line frames in Figs. 3(a)-(e). Fig. 3(j) is more similar to Fig. 3(f) in comparison to Figs. 3(g), (h) and (i). Deblur is achieved to a certain extent in Figs. 3(h) and (i), but the small blood vessels get over-sharpened (as indicated by the green arrows in Figs. 3(h) and (i)) and the boundary between the object and the background appears to have ghost patterns (as indicated by the blue arrows in Figs. 3(h) and (i)).

Quantitative analysis using the above three methods on Test Data are shown in Fig. 4. In Figs. 4(a), (b) and (c), the horizontal axis represents the $\sigma$ value, and the vertical axis represents the MSE, PSNR, and SSIM values. The green curve, the blue curve, and the red curve represent the results using our method, Deconv method, and RL method, respectively. From Fig. 4, it can be seen that the image processed by using our method is superior compared to other two methods since all three quantitative indicators are significantly better than the indicators of the other two methods.

*3.2. Simulated Data*

The results of processing Simulated Data are shown in Fig. 5. The first column shows the original images ((a) and (k) have the same pattern, (f) and (p) have the same pattern), the second column shows the blurred images ((b) and (g) are blurred by a Gaussian window with $\sigma=3$, (l) and (q) are blurred by a Gaussian window with $\sigma=6.5$), the third, fourth, and fifth columns are the deblurred images of the second column images by Deconv method, RL method, and our method, respectively. From the images in Fig. 5, it can be seen that even when the test patterns are different from training patterns, our deblurring effect is still better than other methods. Compared to the blind and RL algorithms, images processed by our method are significantly closer to the original images and have clear boundaries for the following three reasons: (1) deep learning method is efficient in feature extraction and strong generalization capabilities [20]; (2) our method employed a new and efficient RDN to extract different image features easily and efficiently as outlined in the methods section; (3) our models trained the network fully and effectively with no over-fitting phenomenon as shown in the Fig. B.$_1$ (loss curve). Thus, despite simulated pattern being different from the training images, our deep learning method can extract the general features hidden in the training sets and has good generalization ability for other types of data. These results are consistent with existing research, where researchers have used natural images to train the data and then use them to test medical images [38].

Quantitative analyses (QA) of the patterns were performed using MSE, PSNR, and SSIM and results are shown in Table 2. The indicators $QA_{ac}$, $QA_{ad}$, and $QA_{ae}$ in Table 2(a) imply quantitative analyses between Fig. 5(a) and corresponding Figs. 5(c), (d), and(e) respectively; similarly, the indicators $QA_{hf}$, $QA_{if}$, and $QA_{jf}$ in Table 2(b) imply quantitative analyses between Fig. 5(f) and corresponding Figs. 5(h), (i), and(j), respectively; the indicators $QA_{mk}$, $QA_{nk}$, and $QA_{ok}$ in Table 2(c) imply quantitative analyses between Fig. 5(k) and corresponding Figs. 5(m), (n), and(o)), respectively; the indicators $QA_{rp}$, $QA_{sp}$, and $QA_{tp}$ in Table 2(d) imply quantitative analyses between Fig. 5(p) and corresponding Figs. 5(r), (s), and (t) respectively. In this table, the best-performing result in each group is marked in red. As can be seen, the best performing three indicators in this table belongs to the proposed method in this study and are better than the other two methods for any data set.

The signal intensity along the green dotted line in Fig. 5(a), (b), (c), (d), and (e) are plotted in Fig. 6(a) as pink dotted line, gray dotted line, green solid line, red solid line, and blue solid line, respectively. Corresponding intensity line belonging to Figs. 5(f), (g), (h), (i), and (j); (k), (l), (m), (n), and (o); (p), (q), (r), (s), and (t), are plotted in Figs. 6(c), (e) and (g) respectively. To observe the deblurring results at the boundary, the enlarged area of the red dotted box in Figs. 6(a), (c), (e), and (g) are shown in Figs. 6(b), (d), (f), and (h), respectively. It can be seen that the blue solid line is more in sync with the pink dotted line in all the figures. In the enlarged Figs. 6(b) and (d), our method shows a perfect deblurred result at the boundary (the red dotted line and the blue solid line completely coincide at the boundary position), while the other two methods have limited deblurring ability. Even though patterns in Figs. 5(l) and (q) are heavily blurred; our method could deblur the image as shown in Figs. 6(f) and (h), and the deblur effect is more pronounced in our method compared to the other two methods.

*3.3. Experimental Data*

The microscopic images of the rat's eyeball obtained using $10\times$ and $20\times$ objectives are shown in Figs. 7(a) and (c), respectively, and the deblurred image of $10\times$ objective using our method is shown in Fig. 7(b). The Figs. 7(d), (g), and (j); (e), (h), and (k); and (f), (i), and (l) are the enlarged images of three same areas in Figs. 7(a), (b), and (c), respectively. These areas are pointed out in Fig. 7(a) using red solid frames (named 1, 2, and 3). The deblurred image in Fig. 7(b) obtained using our method is similar to Fig. 7(c), which can be seen clearly in the enlarged images. An intensity discrepancy exists between Figs. 7(a) and (c) due to non-ideal imaging conditions between these two acquisitions (in addition to different PSF in different experiments, other conditions will also be different, such as the light intensity, sample placement, focal length, etc.). This difference is also reflected in Fig. 7(b).

Figs. 8(a) and (b) are the original and deblurred depth-encoded (different colors represent different depths) maximum amplitude projection (MAP) vascular images of the rat's eyeball, respectively; Figs. 8(c) and (d) are the enlarged view of the subareas in Figs. 8(a) and (b), respectively. The subareas are indicated by the red dotted frame in Fig. 8(a). From this Fig. 8, it can be seen that the image quality in the deblurred Fig. 8(d) is improved compared to the original Fig. 8(c). The signal intensity values of the blue and red color lines in Figs. 8(c) and (d), respectively, are shown with corresponding colors in Fig. 8(e). Comparing blue and red curves in Fig. 8(e), the advantages of the deblurred image using our method are as follows: (1) significantly improved vascular signal intensity; (2) improved details of the dense vascular plexus (as shown by two green dashed boxes in Fig. 8(e)); (3) enhanced resolution of micro-vessels ($\approx 6\mu m$, as indicated by green arrows in Fig. 8(e) along with enlarged orange box). The effect of deblurring is not so obvious in large blood vessel because large blood vessels have less detail information compared to small blood vessels and the system resolution of $6.7\mu m$ is sufficient to resolve large blood vessels ($\geq 30\mu m$).

To further verify the deblurring ability of our method in OR-PAM, the edge of a sharp metallic blade was imaged using our custom-built OR-PAM system and deblurred with our method. To remove random noise and stabilize the signal, we averaged the signal over 100 acquisitions. Figs. 9(a) and (b) plot the photoacoustic signal amplitude (green cross) as a function of the lateral distance across the edge of the original data and deblurred data, respectively. The line spread function (LSF) in the scanning direction was derived from the first-order derivative. The edge spread function (ESF) and full width at half maximum (FWHM) resolution from this figure were estimated to be $6.73\mu m$ and $3.15\mu m$, respectively, which shows that our method improves the resolution of this system by 2.14× times. This performance is better than the previously proposed deconvolution method [9], which improved the resolution of this system by only 1.9× times.

**4. CONCLUSION**

This is the first time a deep-learning-based deblurring method has been proposed, generally applicable to kinds of optical microscopic imaging systems. The deep learning method does not need any up-sampling (or down-sampling) operations or experimental data during the training process to train models. During deblurring, it only needs input data to be reconstructed into the well-trained CNN to generate deblurred output. Our deblurring is ultra-fast and takes only 0.8 seconds on average, which is independent of hardware or post-processing. We showed the deblurring ability of the proposed method in a variety of data set (Test Data, Simulated Data, and two types of Experimental Data). All our results show that our method not only overcomes the limitations of previously proposed deep learning methods which have special requirements for imaging equipment but also has many advantages compared to other deconvolution methods.

To conclude, the new deep-learning-based deblurring method proposed in this study has the advantages of simple operation, short time-consuming, wide application range, and good deblurring capability. Since the image quality is improved significantly in all types of data used in this study, we expect our method to open up a new path for deblurring not only in optical and photoacoustic microscope but also in all kinds of optical microscopic imaging systems. In addition, our method could also replace traditional deconvolution algorithms and become an algorithm of choice in various biomedical imaging systems.

**Appendices**

*A. The detailed description of the blur operation*

(The relationship between the original object $O(x, y)$ and the output image $I(x, y)$ of the system can be represented as:

$$I(x, y) = O(x, y) * G(x, y) + n(x, y) \tag{A.1}$$

where * denotes the convolution operator in a 2-D plane, $G(x, y)$ denotes the PSF of the system, and $n(x, y)$ denotes the random spatial distribution of noise. Since the optical microscopic system images have a high signal-to-noise ratio (SNR), the effects of noise can be ignored. The traditional optical microscopic imaging system's PSF can be approximated to a Gaussian function:

$$G(x, y) = \frac{1}{2\pi\sigma^2} e^{-(x^2+y^2)/2\sigma^2} \tag{A.2}$$

where $\sigma$ is the variance of the function, which controls the radial range of the function. Different types of convolution operations are simulated by using different sets of $\sigma$.

*B. Loss curves*

Fig. B.1 shows curves of the training loss (orange line) and the verification loss (blue line) of some models ($\sigma = 0.5, 1, 1.5, 2, 2.5, 3, 3.5, 4, 4.5$, and 5, respectively). It can be seen that both the training loss curve and the validation loss curve converge within the training epochs, which means our models have been fully and effectively trained and no over-fitting (or under-fitting) takes place. Loss curves in

Fig. B.₁ are only valid for the data used in this paper. For other data, effectively training without over-fitting (or under-fitting) can be ensured by appropriately increasing or decreasing training data and adjusting the Network's size.

*C. The rules for selecting the model for input data*

The only variable we get for different models is the different $\sigma$ in Eq. (A.₂), which is used to simulate different Gaussian spots in different systems. Thus, we choose the corresponding model by calculating the $\sigma$ in different systems. This problem can be solved by extracting full width at half maximum (FWHM) of the PSF, as the relationship between FWHM and $\sigma$ can be calculated by upgrading Eq. (A.₂) using two points $(0,0)$ and $(x_0, y_0)$：

$$\begin{cases} G(x_0, y_0) = \dfrac{1}{2\pi\sigma^2} e^{-(x_0^2 + y_0^2)/2\sigma^2} \\ G(0,0) = \dfrac{1}{2\pi\sigma^2} \end{cases} \quad (C._1)$$

where $(x_0, y_0)$ is the location at FWHM, $G(0,0) = 2G(x_0, y_0)$ and $FWHM = 2\sqrt{x_0^2 + y_0^2}$, which amounts to $FWHM = 2.355 \times \sigma$.

Many methods are employed to extract the FWHM; we propose two simple methods as follows:
1. FWHM of the PSF can be calculated by:

$$FWHM = 0.51 \dfrac{\lambda}{NA} \quad (C._2)$$

where λ denotes the laser wavelength, and NA denotes the numerical aperture of the optical illumination.
2. FWHM of the PSF can be experimentally measured either by the edge of a sharp metallic blade or ideal particle or resolution test target (Fig. 9 shows this method by measuring the edge of a sharp metallic blade).

**Funding**

This work was supported by the following funds: National Natural Science Foundation of China (NSFC) grants 91739117 and 31570952

**Conflict of Interest**

The authors declare that there are no conflicts of interest

https://doi.org/10.1117/1.JBO.23.3.036015

**Figures**

**FIG. 1**

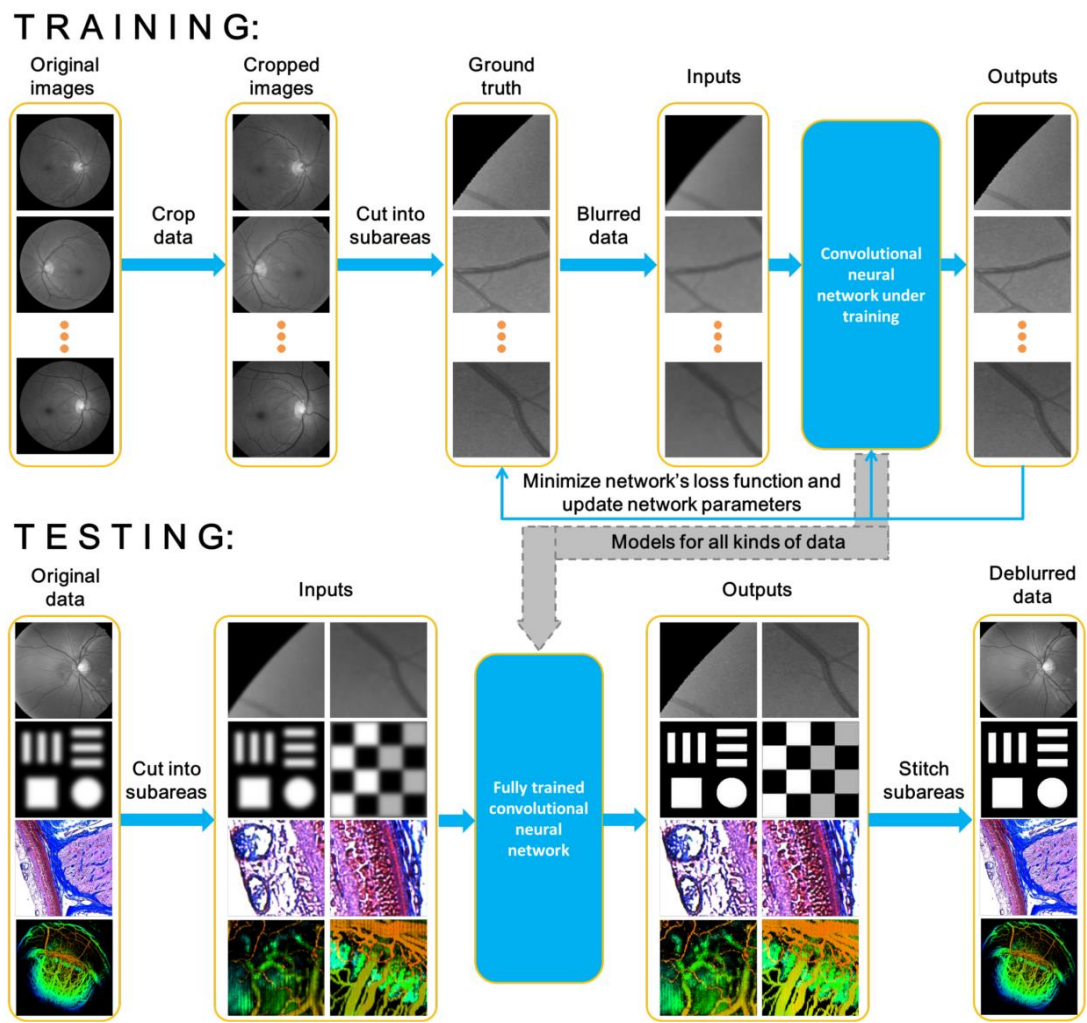

**Fig. 1.** The main framework of our deblurring method. For the TRAINING part, original images (242 ×2912 ×2912 pixels) were downloaded from the database FIRE (https://www.ics.forth.gr/cvrl/fire/), cropped images (242 ×2304 ×2304 pixels) were obtained from original images, ground truth images (19,602 ×256 ×256 pixels) were subareas of cropped images, inputs (19,602 ×256 ×256 pixels) were blurred data of ground truth, finally, outputs are obtained by CNN. For TESTING, original

data were cut into different types of subareas as inputs (n ×256 ×256 pixels), outputs are obtained by inputting these data to well-trained CNN, and deblurred data are obtained by stitching these subareas.

**FIG. 2**

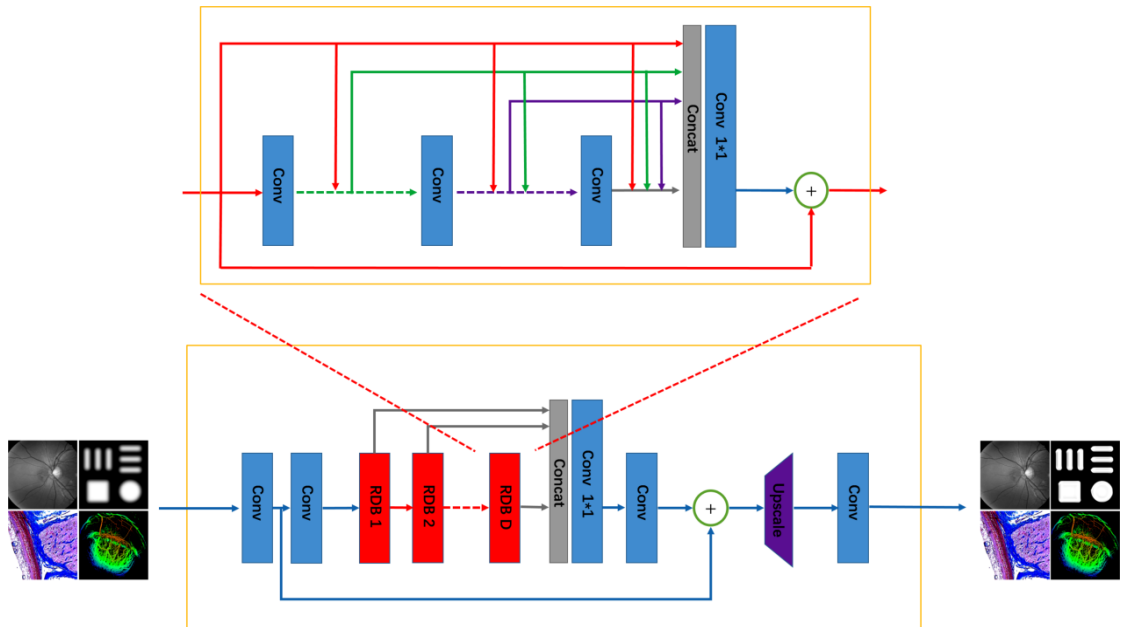

**Fig. 2.** The residual dense networks for optical image deblurring

**FIG. 3**

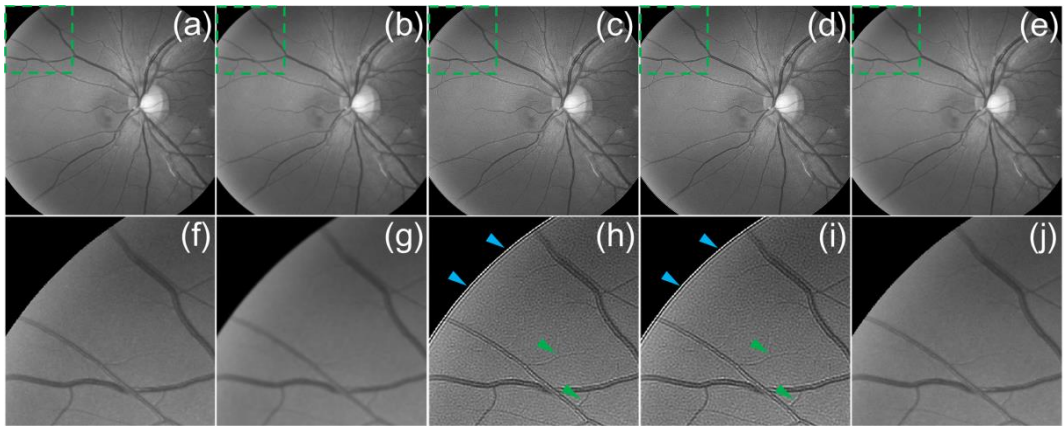

**Fig. 3.** (a) The original image. (b) The blurred image. (c), (d) and (e) The deblurred images processed by Deconv method, RL method, and our method. (f), (g), (h), (i), and (j) The enlarged images at the green dotted line frames of (a), (b), (c), (d), and (e), respectively.

**FIG. 4**

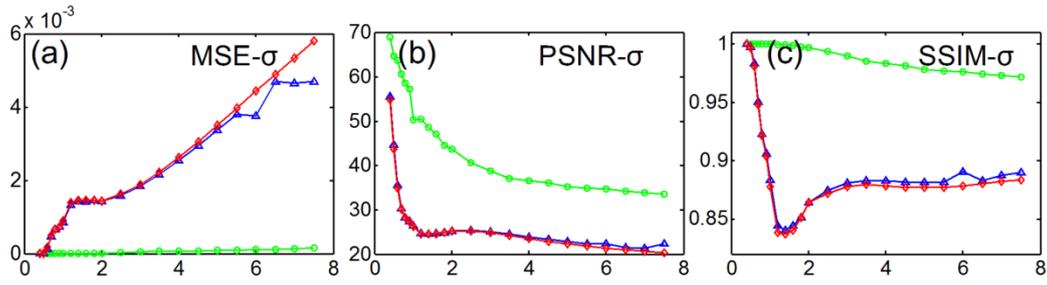

**Fig. 4.** Quantitative analysis of these three different methods of Test Data. The green curve, the blue curve, and the red curve represent the results our method, Deconv method, and RL method.

**FIG. 5**

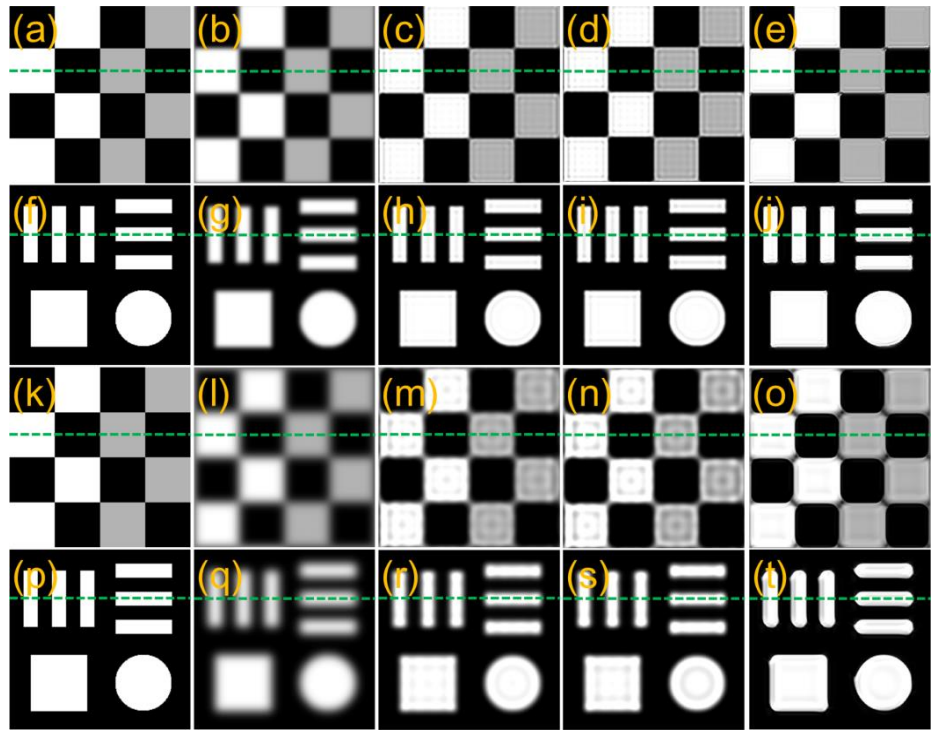

**Fig. 5.** Series of processes for Simulated Data. (a),(f), (k), and (p) are original images, (a) and (k) is the same image, (f) and (p) are the same image. (b) and (g) are blurred images of (a) and (f) by a Gaussian spot with an $\sigma=3$. (l) and (q) are blurred images of (k) and (p) by a Gaussian spot with an $\sigma=6.5$. (c), (d), and (e) are deblurred (b) processed by Deconv method, RL method, and our method, respectively. (h), (i), and (j) are deblurred (f) processed by Deconv method, RL method, and our method, respectively. (m), (n), and (o) are deblurred (k) processed by Deconv method, RL method, and our method, respectively. (r), (s), and (t) is deblurred (p) processed by Deconv method, RL method, and our method, respectively.

**FIG. 6**

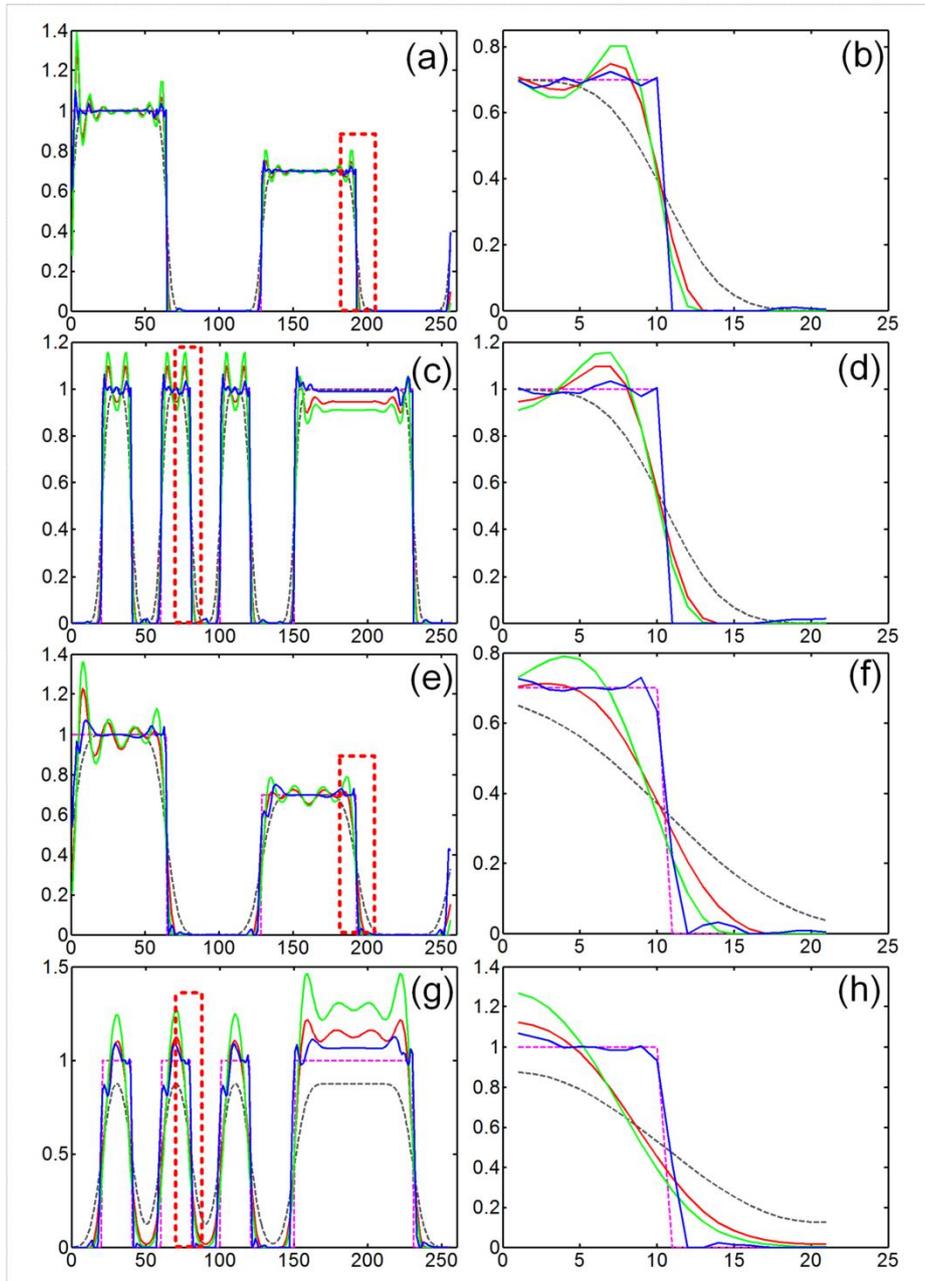

**Fig. 6.** (a) The signal intensity across the green dotted line in the first row of Fig. 5. (c) The signal intensity across the green dotted line in the second row of Fig. 5. (e) The signal intensity of the part corresponding to the green dotted line in the third row of Fig. 5. (g) The signal intensity of the part corresponding to the green dotted line in the last row of Fig. 5. (b), (d), (f), and (h) are the enlarged images of the light blue dotted frame of (a), (c), (e), and (g). Pink dotted lines, gray dotted lines, green solid lines, red solid lines, and blue solid lines represent signal intensities across the green dotted line from first to fifth columns, respectively.

**FIG. 7**

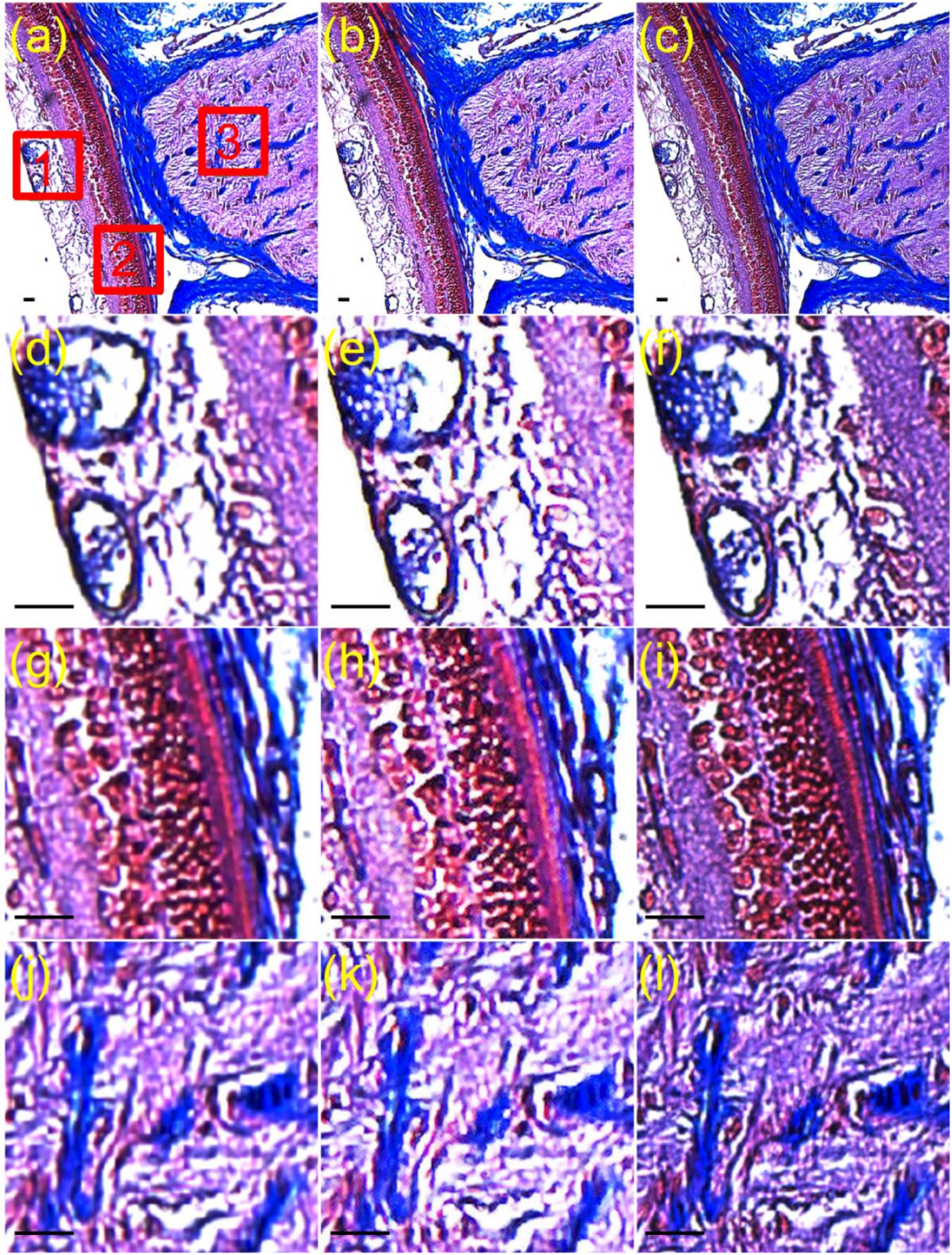

**Fig. 7.** (a), (b) and (c) is the image obtained by the 10× objective, deblurred image of (a) by our method, and the image obtained by the 20× objective, respectively. (d), (g), and (j); (e), (h), and (k); and (f), (i), and (l) are the enlarged images of three same areas in (a), (b), and (c), respectively. These areas are pointed out in (a) using red solid frames (named 1, 2, and 3). Scale bar = 20μm

**FIG. 8**

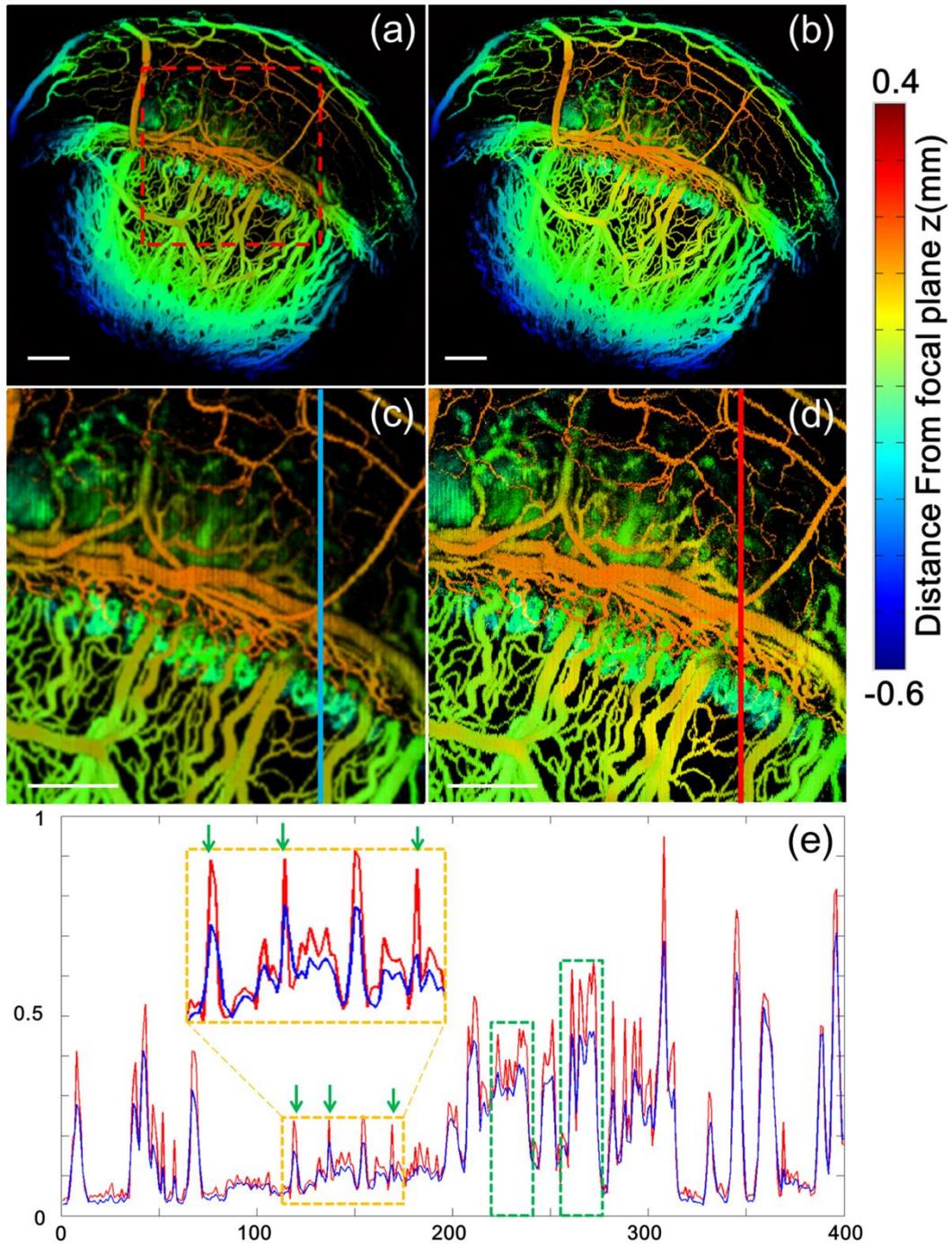

**Fig. 8.** (a) and (b) are the original and deblurred depth-encoded maximum amplitude projection (MAP) rats eye images, respectively. (c) and (d) are the enlarged view of the subareas in (a) and (b), respectively. The subareas are indicated by the red dotted frame in (a). (e) is the signal intensity image of dotted lines in (c) and (d), the blue line is corresponding to the intensity of the blue dotted line in (c), the red line is corresponding to the intensity of the red dotted line in (d). Scale bar $=500 \mu m$.

**FIG. 9**

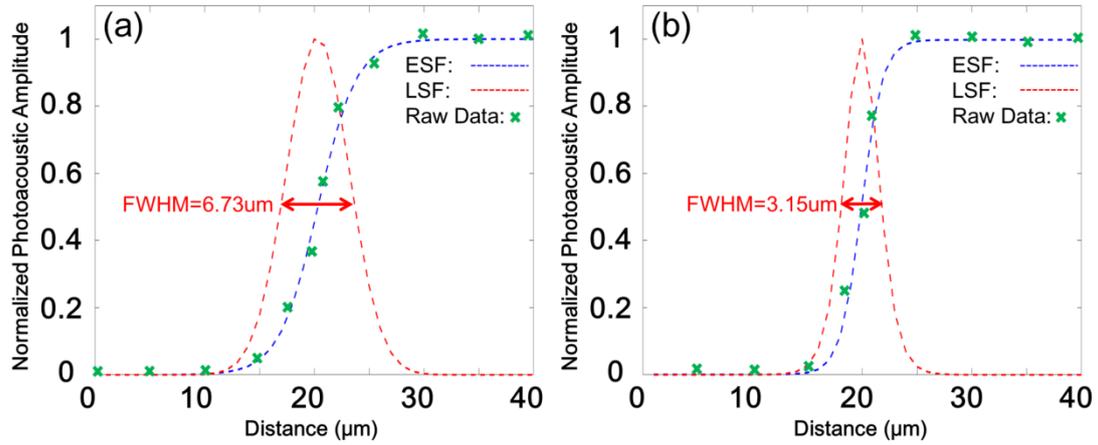

**Fig. 9.** (a) Estimating the full width at half maximum (FWHM) of the original data using the edge of a sharp metallic blade. (b) Estimating the full width at half maximum (FWHM) of the deblurred data. Green cross: original photoacoustic signal; blue dash line: edge spread function (ESF); red dash line: the first-order derivative of the ESF, representing the LSF along the scanning direction.

**FIG. B.1**

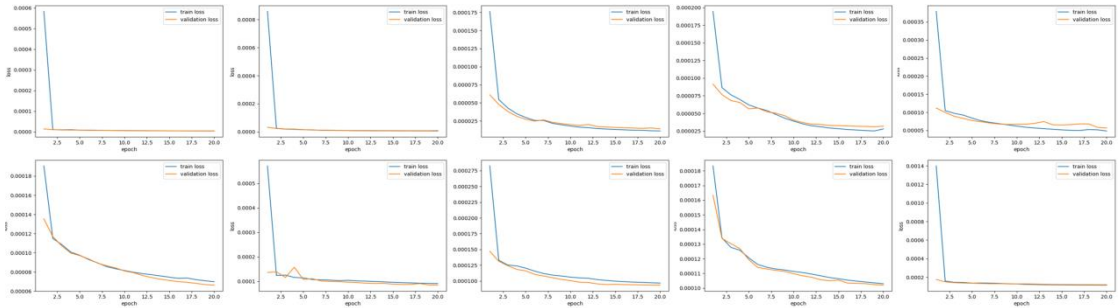

**Fig. B.1.** Loss curves of training and verification data of some models ($\sigma$ = 0.5, 1, 1.5, 2, 2.5, 3, 3.5, 4, 4.5, and 5, respectively)

**Tables**

**Table 1**

| Name | Kernel number | Kernel size | Weights initialization | Activation function |
|---|---|---|---|---|
| "Conv" | 32 | 3*3 | He initialization [31] | ReLU [32] |
| "Conv 1*1" | 32 | 1*1 | He initialization | None |

**Table 1.** The details of the convolutional layers in the RDN.

**Table 2**

| (a) | $QA_{ac}$ | $QA_{ad}$ | $QA_{ae}$ | (b) | $QA_{hf}$ | $QA_{if}$ | $QA_{jf}$ |
|---|---|---|---|---|---|---|---|
| MSE | 0.0079 | 0.0086 | 0.0051 | MSE | 0.0092 | 0.0102 | 0.0033 |
| PSNR | 21.039 | 20.660 | 22.911 | PSNR | 20.352 | 19.927 | 24.820 |
| SSIM | 0.8507 | 0.8209 | 0.8699 | SSIM | 0.8423 | 0.8275 | 0.8796 |
| (c) | $QA_{mk}$ | $QA_{nk}$ | $QA_{ok}$ | (d) | $QA_{rp}$ | $QA_{sp}$ | $QA_{tp}$ |
| MSE | 0.0190 | 0.0200 | 0.0152 | MSE | 0.0223 | 0.0253 | 0.0138 |
| PSNR | 17.207 | 17.000 | 18.189 | PSNR | 16.516 | 15.966 | 18.587 |
| SSIM | 0.7145 | 0.6975 | 0.7598 | SSIM | 0.6666 | 0.6843 | 0.7400 |

**Table 2.** Quantitative comparison between for three different methods for Simulated Data. The best-performing value in each group is marked as red.